\documentclass[prl,amsfonts,showpacs,showkeys,preprint]{revtex4}
\usepackage{graphicx}
\usepackage{eepic}
\usepackage{xcolor}

\RequirePackage{times}
\RequirePackage{mathptm}

\begin{document}


\title{Proposed direct test of a certain type of noncontextuality in quantum mechanics}

\author{Karl Svozil}
\email{svozil@tuwien.ac.at}
\homepage{http://tph.tuwien.ac.at/~svozil}
\affiliation{Institut f\"ur Theoretische Physik, Vienna University of Technology,  \\
Wiedner Hauptstra\ss e 8-10/136, A-1040 Vienna, Austria}

\begin{abstract}
The noncontextuality of quantum mechanics can be directly tested by  measuring two entangled particles with more than two outcomes per particle. The two associated contexts are ``interlinked'' by common observables.
\end{abstract}


\maketitle

Quantum value indefiniteness~\cite{peres222} refers to the impossibility of a consistent coexistence of certain complementary, operationally incompatible quantum observables.
It is inferred from three sources:
(i) from quantum violations of constraints on classical probability distributions termed {\em `conditions of possible experience'} by Boole~\cite{Boole-62},
also known as the Boole-Bell type inequalities~\cite{Pit-94},
(ii) from the Kochen-Specker theorem~\cite{kochen1,svozil-tkadlec,cabello-96}, as well as
(iii) from the Greenberger-Horne-Zeilinger~\cite{ghz,mermin-93} theorem.
Formally, these results are related to the ``scarcity'' or even total absence
of two-valued states identifiable as (classical) truth assignments on the entire range of quantum observables.
In what follows, quantum contextuality~\cite{bohr-1949,bell-66,hey-red,redhead,svozil-2008-ql}
will be identified with the assertion that the result of a measurement depends on what other observables are comeasured alongside of it.
It is one conceivable (but not necessary~\cite{svozil-2006-omni}) quasi-classical interpretation of quantum value indefiniteness,
thereby counterfactually maintaining the ``physical existence'' of the full domain of possible physical observables.

There exist other notions of contextuality based upon violations of some bounds on or conditions imposed by, classical probabilities.
In their extreme form, these amount to all-or-nothing–type contradictions between noncontextual hidden variables and quantum mechanics.
The corresponding experimental tests indicate the occurrence of this type of
quantum contextuality~\cite{cabello-98,Mi-Wein-Zu-2000,Si-Zu-Wein-Ze-2000,huang-2003,hasegawa:230401,cabello:210401,cour:012102,kirch-09,Bartosik-09}.
These findings utilize subsequent measurements of quantum observables contributing to a contradiction with their classical counterparts,
but they have no direct bearing on the experiments proposed here which aim at testing another, more direct form of quantum contextuality.

A quantum mechanical context~\cite{svozil-2008-ql}
is a ``maximal collection of comeasurable observables'' within the nondistributive structure of quantum propositions.
It can be formalized by a single  ``maximal'' self-adjoint operator, such that
every collection of mutually compatible comeasurable operators (such as projections corresponding to yes--no propositions)
are functions thereof~\cite[\S~84]{halmos-vs}.

Different contexts can be {\em interlinked} at one or more common observable(s) whose Hilbert space representation is
identical and independent of the contexts they belong to.
The context independence of the representation of observables by operators (e.g., projectors) in Hilbert space
suggests that quantum contextuality, if it exists, manifests itself in random and uncontrollable single-particle outcomes.
A necessary condition for the interlinking of two or more contexts
by link observable(s) is the requirement that the dimensionality of the Hilbert space must exceed two,
since for lower dimensional Hilbert spaces
the maximal operators ``decay'' into separate, isolated ``trivial'' Boolean sublogics without any common observable.
This is also the reason for similar dimensional conditions on the theorems by Gleason, as well as by Kochen and Specker.


In what follows we propose an experiment capable of directly testing the contextuality hypothesis  via counterfactual elements of physical reality.
Indeed, counterfactual reasoning might be considered less desirable than direct measurements,
as it involves an additional logical inference step rather than a straight empirical finding.

In the proposed experiment, two different contexts or, equivalently, two noncommuting maximal observables,
are simultaneously measured on a pair of spin one particles in a singlet state~\cite{hey-red,stairs83,brown90}.
The contexts are fine-tuned to allow a common {\em single observable interlinking} them.
Although the proposal possesses some conceptual similarities to Einstein-Podolsky-Rosen type experiments, the quantum states
as well as the structure of the observables are different.

We shall first consider the contexts originally proposed by Kochen and Specker~\cite[pp.~71-73]{kochen1},
referring to the change in the energy of the lowest orbital state of orthohelium
resulting from the application of a small electric field with rhombic symmetry.
The terms {\em Kochen-Specker contexts} and (maximal) {\em Kochen-Specker operators} will be used synonymously.
More explicitly, the maximal Kochen-Specker operators associated with this link configuration
can be constructed from the spin one observables (e.g., Refs.~\cite{schiff-55,rose})
in arbitrary directions measured in spherical coordinates
\begin{equation}
\label{l-soksp}
J(\theta , \phi )=
\left(
\begin{array}{cccc}
\cos \theta & {e^{-i\phi}\sin \theta \over \sqrt{2}}& 0      \\
{e^{i\phi}\sin \theta \over \sqrt{2}}& 0
& {e^{-i\phi}\sin \theta \over \sqrt{2}}      \\
0& {e^{i\phi}\sin \theta \over \sqrt{2}}& -\cos \theta
\end{array}\right),
\end{equation}
where  $0 \le \theta \le \pi$ stands for the polar angle in the $x$-$z$-plane taken
from the $z$-axis,
and $0 \le \varphi < 2 \pi$  is the azimuthal angle in the $x$-$y$-plane taken
from the $x$-axis.
The orthonormalized
eigenvectors associated with the eigenvalues $+1$, $0$, $-1$ of
$J(\theta , \phi )$ in Eq.~(\ref{l-soksp})
are
\begin{equation}
\label{l-soksp-ev}
\begin{array}{cccc}
x_{+1}&=e^{i\delta_{+1}}& \left(
e^{-i\phi} \cos^2{\theta \over 2}, {1\over \sqrt{2}}   \sin \theta ,e^{i\phi}  \sin^2{\theta \over 2}
\right),\\
x_{0}&=e^{i\delta_0}& \left(
-{1\over \sqrt{2}} e^{-i\phi} \sin \theta , \cos \theta , {1\over \sqrt{2}} e^{i\phi}\sin \theta
\right),\\
x_{-1}&=e^{i\delta_{-1}}& \left(
e^{-i\phi} \sin^2{\theta \over 2}, - {1\over \sqrt{2}}     \sin \theta , e^{i\phi}\cos^2{\theta \over 2}
\right) ,
\end{array}
\end{equation}
where $\delta_{\pm 1}$, and $\delta_0$ stand for arbitrary phases.

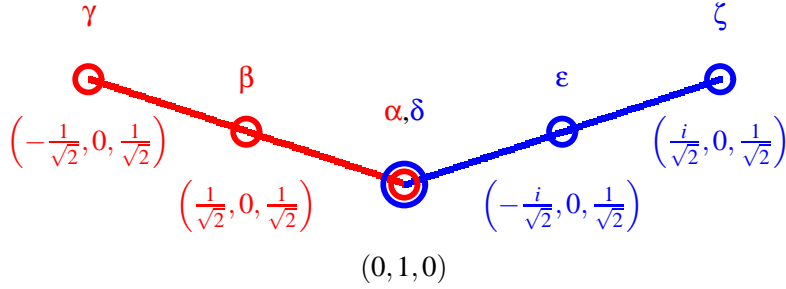
\begin{figure}
\begin{center}
\unitlength1.4mm
\allinethickness{2.2pt} 
\begin{picture}(61.33,36.00)
\multiput(0.33,20.00)(0.36,-0.12){84}{{\color{red}\line(1,0){0.36}}}
\multiput(30.33,10.00)(0.36,0.12){84}{{\color{blue}\line(1,0){0.36}}}
\put(0.33,20.00){{\color{red}\circle{2.50}}}
\put(15.33,15.00){{\color{red}\circle{2.50}}}
\put(30.33,10.00){{\color{red}\circle{2.50}} }
\put(30.33,10.00){{\color{blue}\circle{4.00}} }
\put(45.33,15.00){{\color{blue}\circle{2.50}} }
\put(60.33,20.00){{\color{blue}\circle{2.50}} }
\put(30.33,2.00){\makebox(0,0)[cc]{$(0, 1, 0)$}}
\put(30.33,17.00){\makebox(0,0)[cc]{{\color{red}$\alpha$},{\color{blue}$\delta$}}}
\put(0.33,14.00){\makebox(0,0)[cc]{{\color{red}$\left(-\frac{1}{\sqrt{2}}, 0, \frac{1}{\sqrt{2}}\right)$}}}
\put(0.33,26.00){\makebox(0,0)[cc]{{\color{red}$\gamma$}}}
\put(15.33,8.00){\makebox(0,0)[cc]{{\color{red}$\left(\frac{1}{\sqrt{2}}, 0, \frac{1}{\sqrt{2}}\right)$}}}
\put(15.33,20.00){\makebox(0,0)[cc]{{\color{red}$\beta$}}}
\put(45.33,8.00){\makebox(0,0)[cc]{{\color{blue}$\left(-\frac{i}{\sqrt{2}}, 0, \frac{1}{\sqrt{2}}\right)$}}}
\put(45.33,20.00){\makebox(0,0)[cc]{{\color{blue}$\epsilon$}}}
\put(60.33,14.00){\makebox(0,0)[cc]{{\color{blue}$\left(\frac{i}{\sqrt{2}}, 0, \frac{1}{\sqrt{2}}\right)$}}}
\put(60.33,26.00){\makebox(0,0)[cc]{{\color{blue}$\zeta$}}}
\end{picture}
\end{center}
\caption{(Color online) Diagrammatical representation of two interlinked Kochen-Specker contexts:
 Greechie (orthogonality) diagram representing two tripods with a common leg: points stand for individual basis vectors, and entire contexts
--- in this case the one-dimensional linear subspaces spanned by the vectors of the orthogonal tripods
--- are drawn as smooth curves.
\label{2009-context-f1}}
\end{figure}

For real $\alpha \neq \beta \neq \gamma \neq \alpha$,
the maximal Kochen and Specker operators~\cite{kochen1} are defined by
\begin{equation}
\begin{array}{ccl}
C_{KS}(\alpha , \beta ,\gamma) &=&  \frac{1}{2}\left[ (\alpha  + \beta - \gamma )J^2(\frac{\pi}{2},0) + (\alpha  -  \beta + \gamma) J^2(\frac{\pi}{2},\frac{\pi}{2}) + (\beta + \gamma -\alpha ) J^2(0,0)\right],  \\
C_{KS}' (\alpha , \beta ,\gamma)&=&  \frac{1}{2}\left[ (\alpha  + \beta - \gamma )J^2(\frac{\pi}{2},\frac{\pi}{4}) + (\alpha  -  \beta + \gamma) J^2(\frac{\pi}{2},\frac{3\pi}{4}) + (\beta + \gamma -\alpha ) J^2(0,0)\right].\end{array}
\end{equation}
Their common spectrum of eigenvalues is $\alpha$, $\beta$ and $\gamma$, corresponding to the eigenvectors
$(0, 1, 0)$, $(1, 0, 1)$, $(-1, 0, 1)$ of $C_{KS}$,  and
$(0, 1, 0)$, $(-i, 0, 1)$, $(i, 0, 1)$ of $C_{KS}'$, respectively.
The resulting orthogonality structure of propositions is depicted in Fig.~\ref{2009-context-f1}.

In order to be able to use the type of counterfactual inference employed by an Einstein-Podolsky-Rosen setup,
a multipartite quantum state has to be chosen which satisfies the {\em uniqueness property}~\cite{svozil-2006-uniquenessprinciple}
with respect to the two Kochen-Specker contexts
such that knowledge
of a measurement outcome of one particle entails the certainty that, if this observable were measured on the
other  particle(s) as well, the outcome of the measurement would be a unique function of the
outcome of the measurement actually performed.
Consider the two spin-one particle singlet state
$\left|  \left. \varphi_s \right\rangle  \right. =(1/\sqrt{3})\left(-|00\rangle+|-+\rangle+|+-\rangle\right)$
and identify with the spin states the directions in Hilbert space according to Eqs.~(\ref{l-soksp-ev}); i.e., with
$|+\rangle =(1,0,0)$,
$|0\rangle =(0,1,0)$, and
$|-\rangle =(0,0,1)$; hence in the Kronecker product representation,
$\left|  \left. \varphi_s \right\rangle  \right. =(1/\sqrt{3})\left(0,0,1,0,-1,0,1,0,0\right)$.
This singlet state is form invariant under spatial rotations (but not under all unitary transformations~\cite{rose})
and satisfies the uniqueness property (see below),
just as the ordinary Bell singlet state of two spin one-half quanta (we cannot use these because they are limited to $2 \times 2$ dimensions, with merely two dimensions per quantum).
Hence, it is possible to employ a similar counterfactual argument and establish two elements of physical reality
according to the Einstein-Podolsky-Rosen criterion for
the two interlinked Kochen-Specker contexts $C_{KS}$ as well as $C_{KS}'$.

When combined with the singlet state  $\left|  \left. \varphi_s \right\rangle  \right.$,
two ``collinear'' Kochen-Specker contexts yield
\begin{equation}
\begin{array}{l}
\text{Tr}\left\{ \bigl| \varphi_s \right\rangle \left\langle \varphi_s  \bigr|
\;\cdot \;
\left[C_{KS}(\alpha , \beta , \gamma)\otimes C_{KS}( \delta ,\epsilon , \zeta )\right]\right\}
= \\
\qquad =
\text{Tr}\left\{ \bigl| \varphi_s \right\rangle \left\langle \varphi_s  \bigr|
\;\cdot \;
\left[C_{KS}'(\alpha , \beta , \gamma)\otimes C_{KS}'( \delta ,\epsilon , \zeta )\right]\right\}
=
\frac{1}{3} \left[ \alpha  \delta + \beta \epsilon+ \gamma \zeta \right]
.
\end{array}
\end{equation}
As a consequence, in this configuration the uniqueness property manifests itself
by the unique joint occurrence of the outcomes associated with $\alpha \leftrightarrow \delta$
(corresponding to the proposition associated with the link observable between $C_{KS}$ and $C_{KS}'$),
as well as $\beta \leftrightarrow \epsilon$  and  $\gamma \leftrightarrow \zeta$.
Thus, by counterfactual inference, if the contexts measured on both sides are identical,
whenever $\alpha$, $\beta$ or $\gamma$ is registered on one side,  $\delta$, $\epsilon$ or $\zeta$
is measured on the other side, respectively, and {\it vice versa}.

We are now in the position to formulate a testable criterion for (non)contextuality:
Contextuality predicts that there exist outcomes associated with $\alpha$ on one context $C_{KS}$ which
are accompanied by the outcomes $\epsilon$ or $\zeta$ for the other context $C_{KS}'$;
likewise $\delta$ should be accompanied by $\beta$ and $\gamma$.
The quantum mechanical expectation values can be obtained from
\begin{equation}
\text{Tr}\left\{ \bigl| \varphi_s \right\rangle \left\langle \varphi_s  \bigr|
\;\cdot \;
\left[C_{KS}(\alpha , \beta , \gamma)\otimes C'_{KS}( \delta ,\epsilon , \zeta )\right]\right\}
=\frac{1}{6} \left[2 \alpha  \delta + (\beta + \gamma) (\epsilon + \zeta) \right]
.
\end{equation}
As a consequence, the outcomes
$\alpha$--$\epsilon $,
$\alpha$--$\zeta  $, as well as
$\beta $--$ \delta $ and
$\gamma $--$ \delta $ indicating contextuality do not occur.
This is in contradiction with the contextuality hypothesis.


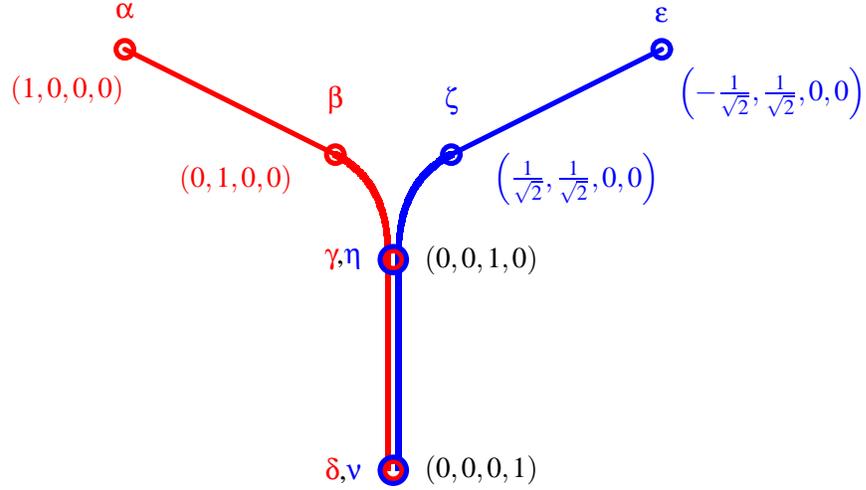
\begin{figure}
\begin{center}
\unitlength 0.70mm
\allinethickness{2pt} 
\begin{picture}(103.67,92.00)
\put(0.00,80.33){{\color{red}\line(2,-1){40.00}}}
{{\color{red} \bezier{104}(40.00,60.33)(50.33,55.00)(50.00,40.33)}}
\put(50.00,40.33){{\color{red}\line(0,-1){40.00}}}
\put(102.00,80.33){{\color{blue}\line(-2,-1){40.00}}}
{{\color{blue} \bezier{104}(62.00,60.33)(51.67,55.00)(52.00,40.33)}}
\put(52.00,40.33){{\color{blue}\line(0,-1){40.00}}}
\put(0.00,80.33){{\color{red}\circle{3.33}}}
\put(40.00,60.33){{\color{red}\circle{3.33}}}
\put(51.00,40.33){{\color{red}\circle{3.33}}}
\put(51.00,40.33){{\color{blue}\circle{5}}}
\put(51.00,0.33){{\color{red}\circle{3.33}}}
\put(51.00,0.33){{\color{blue}\circle{5}}}
\put(102.00,80.33){{\color{blue}\circle{3.33}}}
\put(62.00,60.33){{\color{blue}\circle{3.33}}}
\put(45.00,40.33){\makebox(0,0)[rc]{{\color{red}$\gamma$},{\color{blue}$\eta$}}}
\put(57.00,40.33){{\makebox(0,0)[lc]{$(0,0,1,0)$}}}
\put(45.00,0.33){\makebox(0,0)[rc]{{\color{red}$\delta$},{\color{blue}$\nu$}}}
\put(57.00,0.33){{\makebox(0,0)[lc]{$(0,0,0,1)$}}}
\put(0.00,72.66){{\color{red}\makebox(0,0)[rc]{$(1,0,0,0)$}}}
\put(0.00,87.66){{\color{red}\makebox(0,0)[cc]{$\alpha$}}}
\put(32.00,55.66){{\color{red}\makebox(0,0)[rc]{$(0,1,0,0)$}}}
\put(40.00,70.66){{\color{red}\makebox(0,0)[cc]{$\beta$}}}
\put(70.00,55.66){{\color{blue}\makebox(0,0)[lc]{$\left(\frac{1}{\sqrt{2}},\frac{1}{\sqrt{2}},0,0\right)$}}}
\put(62.00,70.66){{\color{blue}\makebox(0,0)[cc]{$\zeta$}}}
\put(105.00,72.00){{\color{blue}\makebox(0,0)[lc]{$\left(-\frac{1}{\sqrt{2}},\frac{1}{\sqrt{2}},0,0\right)$}}}
\put(102.00,87.00){{\color{blue}\makebox(0,0)[cc]{$\epsilon$}}}
\end{picture}
\end{center}
\caption{(Color online) Greechie diagram of two contexts in four-dimensional Hilbert space interconnected by two link observables.
\label{2009-context-f2}}
\end{figure}
Another context configuration in four-dimensional Hilbert space  drawn in Fig.~\ref{2009-context-f2}
consists of two contexts which are interconnected by
{\em two} common link observables.
The two context operators
\begin{equation}
\label{l-soksp41}
C(\alpha , \beta , \gamma , \delta )=
\text{diag}
\left(
\alpha , \beta , \gamma , \delta
\right),
\qquad
\qquad
C'(\alpha , \beta , \gamma , \delta )=
\text{diag}
\left(
\begin{array}{cccc}
\frac{\alpha + \beta}{2}  & \frac{\alpha  - \beta}{2}   \\
\frac{\alpha  - \beta}{2}  & \frac{\alpha  + \beta}{2}  \\
\end{array}, \gamma , \delta
\right)
\end{equation}
have identical eigenvalue spectra containing
mutually different real eigenvalues  $\alpha$,
$\beta$,
$\gamma$ and $\delta$.

Consider the singlet state of two spin-$3/2$ observables
$\left|  \left. \psi_s \right\rangle  \right. =
\frac{1}{2} \left(
\left| \left. \frac{3}{2}, -\frac{3}{2}\right\rangle \right.
 - \left| \left.  -\frac{3}{2}, \frac{3}{2}\right\rangle    \right.
- \left| \left.  \frac{1}{2}, -\frac{1}{2}\right\rangle  \right.
+ \left| \left.  -\frac{1}{2}, \frac{1}{2}\right\rangle   \right.
\right)
$ satisfying the uniqueness property for all spatial directions.
The four different spin states can be identified with the cartesian basis of fourdimensional Hilbert space
$\left| \left. \frac{3}{2}\right\rangle \right. = (1,0,0,0)$,
$\left| \left. \frac{1}{2}\right\rangle \right. = (0,1,0,0)$,
$\left| \left. -\frac{1}{2}\right\rangle \right. = (0,0,1,0)$,
and
$\left| \left. -\frac{3}{2}\right\rangle \right. = (0,0,0,1)$,
respectively.
When combined with the singlet state  $\left|  \left. \psi_s \right\rangle  \right.$,
two ``collinear'' contexts yield
\begin{equation}
\begin{array}{rcl}
\text{Tr}\left\{ \bigl| \psi_s \right\rangle \left\langle \psi_s  \bigr|
\;\cdot \;
\left[C(\alpha , \beta , \gamma, \delta )\otimes C(  \epsilon , \zeta ,\eta , \nu )\right]\right\}
&=&
\frac{1}{4} \left[ \alpha  \nu + \beta \eta + \gamma \zeta + \delta \epsilon \right],\\
\text{Tr}\left\{ \bigl| \psi_s \right\rangle \left\langle \psi_s  \bigr|
\;\cdot \;
\left[C'(\alpha , \beta , \gamma, \delta )\otimes C'( \epsilon , \zeta ,\eta , \nu )\right]\right\}
&=&
\frac{1}{8} \left[  \left( \alpha  +  \beta \right) \left(\eta  + \nu \right) + \left( \gamma + \delta \right)  \left( \epsilon + \zeta \right) \right]
.
\end{array}
\end{equation}
As a consequence, in this configuration the uniqueness property manifests itself
by the unique joint occurrence of the outcomes associated with $\alpha \leftrightarrow \nu$ and $\beta \leftrightarrow \eta$,
as well as $\gamma  \leftrightarrow \zeta$  and  $\delta \leftrightarrow \epsilon$ for $C$, and
$(\alpha \text{ or } \beta ) \leftrightarrow (\eta \text{ or } \nu )$,
as well as
$(\gamma \text{ or } \delta ) \leftrightarrow (\epsilon \text{ or } \zeta )$
for $C'$.
Thus, by counterfactual inference, if the contexts measured on both sides are identical,
whenever $\alpha$ or $\beta$, and $\gamma$ or $\delta$ is registered on one side, $\nu$ or $\eta$, and  $\zeta$ or $\epsilon$
is measured on the other side, respectively, and {\it vice versa}.

Compared to the previous Kochen-Specker contexts, this configuration has the additional advantage that ---
in the absence of any criterion for outcome preference  ---
Jayne's principle~\cite{jaynes-prob} suggests that
contextuality predicts totally uncorrelated outcomes associated with a maximal unbias of the two common link observables,
resulting in the equal occurrence of the joint outcomes
$\gamma$--$\eta$,
$\gamma$--$\nu$,
$\delta$--$\eta$, and
$\delta$--$\nu$.
The quantum mechanical predictions are based on the expectation values
\begin{equation}
\text{Tr}\left\{ \bigl| \psi_s \right\rangle \left\langle \psi_s  \bigr|
\;\cdot \;
\left[C(\alpha , \beta , \gamma , \delta )\otimes C'(\epsilon , \zeta , \eta , \nu )\right]\right\}
=\frac{1}{8} \left[2 \left( \alpha  \nu + \beta \eta \right)+ \left(\gamma + \delta\right) \left(\epsilon + \zeta \right) \right]
.
\end{equation}
As a consequence, there are no outcomes
$\gamma$--$\eta$,
$\gamma$--$\nu$,
$\delta$--$\eta$, and
$\delta$--$\nu$, which is in contradiction to the contextuality postulate.

One of the conceivable criticisms against the presented arguments is that the configurations considered, although containing complementary contexts,
still allow even a full, separable set of two-valued states,
and therefore need no contextual interpretation.
However, it is exactly these Kochen-Specker type contexts which enter the Kochen-Specker argument.
Hence, they should not be interpreted as separate, isolated sublogics, but as parts of a continuum of sublogics,
containing the finite structure devised by Kochen and Specker and others.

One could also point out that it might suffice to
{\em prepare} the particle in some link state ``along'' one context, and then {\em measure} its state
``along'' a different context  ``containing'' the same link observable.
This could for instance in the three-dimensional configuration
be realized by two successive three-port beam splitters arranged serially.
In such a configuration, if the outcomes of the two beam splitters do not coincide at the link observable,
then noncontextuality is disproved; likewise, if there is a perfect correlation between the link state prepared and the link
observable  measured, then contextuality could be disproved.
This configuration might be criticized by proponents of contextuality as being too restrictive,
since there is a preselection, effectively fixing the
preparation state corresponding to the link observable.

Third, one could reprehend that the entangled particles cannot be thought of as isolated and that the singlet state enforces noncontextuality by the way it is constructed.
This criticism could be counterpointed by noting that it is exactly this kind of configurations which yield violations of Boole-Bell type {\em conditions of physical experience.}

The situation can be summarized as follows.
The direct measurement of more than one context on a single particle is blocked by quantum complementarity.
For the counterfactual ``workaround'' to measure two noncommuting interlinked contexts on pairs of spin-one and spin three-half particles in singlet states,
quantum mechanics predicts noncontextual behavior.
Because of the lack of a uniqueness property, counterfactual inference of configurations with more than two particles are impossible .


\end{document}